\documentclass[prd,amsfonts,amssymb,
amsmath,showkeys,showpacs]{revtex4}

\date{May 10, 2007}
\usepackage{graphicx}

\begin{document}

\author{Xing Huang}
\author{Leonard Parker}
\affiliation{Physics Department, University of Wisconsin-Milwaukee,
P.O.Box 413, Milwaukee, Wisconsin USA 53201}

\title{Graviton Propagator in a Covariant Massive Gravity Theory}

\begin{abstract}
We study the massive gravity theory proposed by Arkani-Hamed, Georgi and Schwartz. In this theory, the graviton becomes massive when general covariance is spontaneously broken through the introduction of a field that links two metrics, one of the which will eventually decouple. The excitation of this ``link" field acts like a Goldstone boson
in giving mass to the graviton. We work out the graviton and Goldstone boson propagators explicitly by means of gauge fixing terms similar to the renormalizability
gauges used in gauge theories. With these propagators, we calculate the lowest order tree-level interaction between two external energy momentum tensors. The result is independent of the gauge parameter, but different from the prediction of massless gravity theory, \textit{i.e.,} general relativity. This difference remains even if the mass of the graviton goes to zero, in which case it gives the van Dam-Veltman-Zakharov 
(vDVZ) discontinuity between the propagators of a massive and massless linearized 
graviton. Moreover, we show that the Fierz-Pauli graviton mass term can be considered 
as the ``unitary gauge'' of a more general theory with an extra scalar field. We 
explicitly construct such a theory in which the vDVZ discontinuity arises with a 
graviton mass term that is different from the Fierz-Pauli mass term. This theory has
a local Weyl symmetry under conformal transformations of the metric. In the case when
the mass goes to zero, the Weyl summetry becomes a global symmetry. It  is
possible that the local Weyl symmetry will give a hint as to the form of the corresponding fully nonlinear theory having a nonzero graviton mass.
%
\end{abstract}
\newpage

\maketitle

\section{Introduction}

The theory of massive gravity has been studied extensively. Modification of the infrared behavior of gravity has been considered 
as an alternative to dark energy. Some of these  models 
(for example \cite{Arkani-Hamed:2003uy}\cite{Deffayet:2001pu}%
\cite{Dvali:2000hr}) have similar properties to massive gravity. \\

However it is not an easy task to formulate a consistent theory of massive gravity that approaches general relativity when the graviton mass goes to 0. One of the major difficulties is the well-known van Dam-Veltman-Zakharov discontinuity \cite{vanDam:1970vg}\cite{Iwa}\cite{Zakharov}. The prediction of massive and massless graviton theories for the bending of light by the Sun are different. These differences will remain even if the mass of the graviton goes to 0. So if this discontinuity is unavoidable, one may rule out massive gravity theory by solar system tests. However one may wonder whether such a discontinuity would disappear in a theory where the graviton obtains mass by a spontaneous symmetry breaking mechanism. This speculation is well founded. In spontaneously broken gauge theory, the massive gauge boson propagator in the unitary gauge also appears not to have a limit in which it approaches the massless propagator since there is a term proportional to $k_\mu k_\nu \over M^2$ that will blow up. Without taking into account the Goldstone boson one would reach the conclusion that a kind of vDVZ discontinuity occurs in Yang-Mills theory\cite{vanDam:1970vg}. If one instead starts from a gauge invariant theory and chooses other gauges where a Goldstone boson and a ghost appear, the gauge boson propagator does have a continuous massless limit. With the introduction of mass to the graviton by the usual Fierz-Pauli approach\cite{Fierz-Pauli}, the linearized gravitational theory loses general covariance under infinitesimal coordinate transformations. One may restore general covariance by realizing \cite{Arkani-Hamed:2002sp} that the Fierz-Pauli gauge is the ``unitary gauge'' of a generally covariant theory that involves an additional vector Goldstone-like field. From the analogy with Yang-Mills theory, one may question whether the vDVZ discontinuity and its disagreement with observation will appear in a generally covariant theory of massive gravity.\\

Vainshtein proposed the first covariant massive gravity theory \cite{Vainshtein:1972sx}. He calculated the spherical metric produced by a source with mass parameter $M$ and found that, in the perturbative calculation based on power expansion in the gravitational constant $G$, his theory does contain the same discontinuity as found by vDVZ.
However, the graviton of mass $m_g$ also becomes strongly coupled at the energy scale $\Lambda = {(\frac {m_g{}^4} {G M})}^{1/5}$ in the presence of the source of mass $M$. Above this energy scale, or equivalently, within the radius $\Lambda^{-1}$ (the Vainshtein radius), he showed that the perturbative expansion in $G$ is not trustworthy. Since the arguments for existence of this discontinuity are all based on perturbative calculations, one can not conclude whether or not there is such a discontinuity in the region where the perturbative expansion breaks down. By using an expansion that is non-analytic in $G$, in Ref. \cite{Deffayet:2001uk} they argued that the full non-linear theory has a continuous limit with no vDVZ discontinuity.\\  

Arkani-Hamed, Georgi and Schwartz \cite{Arkani-Hamed:2002sp} developed another model in which massive gravity theory appears as an effective theory with spontaneously broken general coordinate covariance. This theory involves  two metrics related by a link field whose excitation around its vacuum expectation value can be considered as a Goldstone field. This Goldstone field is a vector field and its scalar mode will have strong coupling behavior. This scalar mode can be related \cite{Arkani-Hamed:2002sp} to the longitudinal mode of the massive graviton in the unitary gauge by the Goldstone boson equivalence theorem. Due to this strong coupling (inversely proportional to $m_g$), there is an extra contribution from the scalar mode in the massless limit even at tree level. So the amplitude of the single massive graviton exchange process in this theory is different from that in the massless theory. \\

Here we calculate the graviton and the Goldstone boson propagators that are not given explicitly in ref.\cite{Arkani-Hamed:2002sp} and confirm their result of the existence of a discontinuity at the linearized level. We expand around a flat background and calculate the graviton propagator by a gauge-fixing method similar to using 
the 't~Hooft $R_\xi$ renormalization gauge in spontaneously 
broken gauge 
theory\cite{Fujikawa:1972fe}. With a proper choice of $R_\xi$ gauge, we show that the mixing between kinetic terms of the graviton and the Goldstone boson can be removed and one can obtain their propagators separately. We also show that the result is 
independent of $\xi$.
During the preparation of this paper, we noticed that 
Nibbelink \textit{et al.} \cite{Nibbelink:2006sz} used a particular 
gauge fixing term to obtain the propagators.
In the present paper, we use a two-parameter family of gauge-fixing terms for which we are able to obtain the explicit propagators.\\

In Section 3, we explicitly construct a linearized theory in which the vDVZ discontinuity arises with a graviton mass term that is different from the Fierz-Pauli mass term. This theory has a new scalar
field that allows us to make the theory conformally invariant at
linear order. Although the graviton propagator in this theory has
a mass term that is not that of Fierz and Pauli, the new
scalar field nevertheless gives rise to the vDVZ discontinuity.
This theory is gauge invariant and in the unitary gauge (in
which the scalar field is absent) reduces to the Fierz-Pauli
theory. This way of introducing a graviton mass term that is
different from Fierz-Pauli and still avoids the problem of ghosts
is new to our knowledge.\\

\section{Graviton Propagator in a Gauge Invariant Massive Gravity Theory}
In the covariant massive gravity theory proposed by Arkani-Hamed \textit{et al.}, one has the action 
\begin{equation}\label{Lag} S_{grav+mass} = \int d^4 x \sqrt{-g}(-{M_{Pl}}^2 R[g])+ \int d^4 x \sqrt{-g}(aH H+bH_{\mu\nu}H^{\mu\nu})\end{equation}
where the second term gives a mass to graviton. Here we use the metric with $(+\ -\ -\ -)$ signature which is \emph{different} from the convention in ref.\cite{Arkani-Hamed:2002sp}. Moreover we have already taken the limit in which the other metric in the bi-metric theory decouples. At linearized level, we have
\begin{equation}H_{\mu\nu}=h_{\mu\nu}+\pi_{\mu,\nu}+\pi_{\nu,\mu},\end{equation}
At quadratic level of $h_{\mu\nu}$, the first term of the Lagrangian is
\begin{eqnarray} \label{2} 
\sqrt{-g}R=-\frac{1}{4}({h_{\mu\nu}}^{,\lambda}{h^{\mu\nu}}_{,\lambda}-h^{,\lambda}h_{,\lambda}-2{h^{\mu\lambda}}_{,\lambda}{h_{\mu\rho}}^{,\rho}+2h^{,\mu}{h_{\mu\lambda}}^{,\lambda}).
\end{eqnarray}
Under an infinitesimal coordinate transformation (gauge transformation), $h_{\mu\nu}$ and the Goldstone field $\pi_{\mu}$ transform as
\begin{equation} \label{gauge} h_{\mu\nu} \to h_{\mu\nu} + \xi_{\mu,\nu}+\xi_{\nu,\mu}, \quad \pi_\mu \to \pi_\mu - \xi_\mu .\end{equation} 
One can see that $H_{\mu\nu}$ is invariant under such a gauge transformation. One can obtain the graviton propagator and study its form in the limit when the mass parameters a,b go to zero. One of the obstacles to get the propagator is the mixing term between $h_{\mu\nu}$ and $\pi_{\mu}$. Of course same kind of mixing occurs in gauge theory, where the mixing is removed by a proper gauge fixing. We are going to do the same thing soon. But first of all, it is convienient to separate $\pi_{\mu}$ as 
\begin{equation}\pi_\mu= A_\mu+\partial_\mu \phi\end{equation}
with the introduction of a new artificial gauge symmetry,
\begin{equation}A_\alpha \to A_\alpha+\partial_\alpha \Lambda,\quad \phi \to \phi - \Lambda \end{equation}
the mass term (second term in the Lagrangian \eqref{Lag}) will give terms like 
\begin{equation}\int d^4 x 4a \phi_{,\mu ,\nu}\phi^{,\mu ,\nu}+ 4b \Box \phi \Box \phi = \int d^4 x (a+b)\Box \phi \Box \phi\end{equation}
One need to have $a+b=0$ to avoid the pathological kinetic term with four derivatives which will lead to tachyon or ghost. This requirement will lead to the Fierz-Pauli mass term\cite{Fierz-Pauli}.
\begin{equation}\label{5}f^4 (h_{\mu\nu}h^{\mu\nu}-h^2)\end{equation}
where $f^4$ is some dimensionful constant and can be defined as the graviton mass,
\begin{equation}-{m_g}^2={f^4 \over {M_{Pl}}^2}\end{equation}
The $A_\mu$ field does have an appropriate kinetic term so we will not consider it in the rest of this paper. Moreover there is no mixing between $A_\mu$ and $\phi$,
\begin{equation}\int d^4 x 4a A_{\mu,\nu}\phi^{,\mu,\nu}+ 4b {A^\mu}_{,\mu} \Box \phi = \int d^4 x (a+b){A^\mu}_{,\mu} \Box \phi =0\end{equation}
But there are mixing terms between $A_\mu$ and $h_{\mu\nu}$,
\begin{equation}f^4 (A_{\mu,\nu}h^{\mu\nu}-h{A^\mu}_{,\mu})\end{equation}
and the mixing terms between $\phi$ and $h_{\mu\nu}$ are,
\begin{equation}\label{4} f^4 (\phi_{,\mu,\nu}h^{\mu\nu}-h\Box \phi)\end{equation}
Now one can try to use the following gauge fixing term to remove the mixing between $A_\mu$ and $h_{\mu\nu}$,
\begin{equation}\label{3a} 1/2(\zeta{M_{Pl}}({h_{\mu\nu}}^{,\nu}-h_{,\mu})+{M_{Pl}{m_g}^2 \over \zeta}A_\mu)^2\end{equation}
(For more information about the gauge fixing and measure see \textit{e.g.} \cite{Fradkin:1974df}.) This gauge fixing term will give the following terms that only contain graviton field $h_{\mu\nu}$,
\begin{equation} \label{3} 1/2\zeta^2{M_{Pl}}^2({h^{\mu\nu}}_{,\nu}-h_{,\mu})^2\end{equation}
However even with this term, the kinetic terms of the graviton \eqref{2} will not be invertible. The reason is that one still has the residual gauge symmetry
\begin{equation}\label{6} h_{\mu\nu} \to h_{\mu\nu} + \partial_\mu \partial_\nu \alpha,\quad \phi \to \phi - 1/2 \alpha \end{equation}
\eqref{3} is invariant under this subset of gauge transformation \eqref{gauge}. However the gauge fixing term for this residual symmetry is not enough to remove the mixing terms between $h_{\mu\nu}$ and $\phi$, so we will keep this symmetry and these terms intact at this moment. Instead one can use the following redefinition 
\begin{equation}\label{redef} h_{\mu\nu}={\widetilde{h}}_{\mu\nu}-(1-\epsilon){m_g}^2\phi \eta_{\mu\nu}\end{equation}
to remove the mixing term \eqref{4} between $h_{\mu\nu}$ and $\phi$. Under such a transformation \eqref{4}, the kinetic part of the graviton \eqref{2} will give the contribution
\begin{equation}f^4 (1-\epsilon)(\phi_{,\mu,\nu}{\widetilde{h}}^{\mu\nu}-{\widetilde{h}}\Box \phi).\end{equation}
(The sign is correct since there is a minus sign in the first term of the definition of Lagrangian \eqref{Lag}.) At the same time, the gauge fixing term \eqref{3} will give 
\begin{equation}6 D (1-\epsilon)f^4(\phi_{,\mu,\nu}{\widetilde{h}}^{\mu\nu}-{\widetilde{h}}\Box \phi)\end{equation}
where $D \equiv \frac{1}{2} \zeta^2$. In order to remove the mixing terms one can set $D={-\epsilon \over 6(1-\epsilon)}$. The kinetic part of the graviton \eqref{2} will give the kinetic term for $\phi$,
\begin{equation}\frac{3}{2}(1-\epsilon)^2 f^4 {m_g}^2\phi_{,\mu}\phi^{,\mu}\end{equation}
and the mixing term \eqref{4} will give,
\begin{equation}-3 (1-\epsilon) f^4 {m_g}^2 \phi_{,\mu}\phi^{,\mu}\end{equation}
and \eqref{3} will give,
\begin{equation}-\frac{3}{2}(1-\epsilon)\epsilon f^4 {m_g}^2\phi_{,\mu}\phi^{,\mu}\end{equation}
Combining these three terms one will have the kinetic term for $\phi$
\begin{equation}\label{scalar-K} -\frac{3}{2}(1-\epsilon) f^4 {m_g}^2\phi_{,\mu}\phi^{,\mu}.\end{equation}
We normalize $\phi$ to get the canonical kinetic term,
\begin{equation}\phi = {1 \over M_{Pl} {m_g}^2} \phi_C\end{equation}
There is also a mixing term generated from \eqref{5} and it is proportional to $\tilde h \phi$. But we have a residual symmetry \eqref{6} and thus can add an extra gauge fixing term with the form proportional to $(\chi \tilde h-\frac \phi \chi)^2$ (where $\chi$ is a gauge parameter we choose) to remove this mixing term between $\tilde h$ and $ \phi$.\\

We can introduce external matter sources characterized by two energy momentum tensors $T_{\mu\nu} = T^{a}_{\mu\nu} + T^{b}_{\mu\nu}$. $T^{a}_{\mu\nu}$ and $T^{b}_{\mu\nu}$ are localized at two different points in the position space. The coupling between source and the metric will introduce a term proportional to $T_{\mu\nu} h^{\mu\nu}$ into the Lagrangian \eqref{Lag}. The redefinition \eqref{redef} will also produce a interaction term between $\phi$ and the energy momentum tensor ${T^{\mu}}_\nu$ given by
\begin{equation}{T^{\mu}}_\nu{h^{\nu}}_\mu={T^{\mu}}_\nu({{\widetilde{h}}^{\nu}}_\mu-{m_g}^2\phi{\delta^{\nu}}_\mu)={T^{\mu}}_\nu{{\widetilde{h}}^{\nu}}_\mu-{1 \over M_{Pl}}(1-\epsilon) \phi_C T ,\end{equation}
where $T = T^{\mu}{}_{\nu} \delta^{\nu}{}_{\mu}$.
After evaluating the path integral (perturbatively) of the action with these source terms, one can read off the interaction between these two sources by looking at those terms with exactly one $T^{a}_{\mu\nu}$ and one $T^{b}_{\mu\nu}$. The $\phi$ will provide an extra contribution to the interaction between two sources and this contribution will not go away as $m_g$ goes to zero. The extra contribution from this scalar mode in the massless limit is 
\begin{equation}\label{7} (1-\epsilon)T^{a} T^{b} \over 6 p^2 \end{equation}
When the graviton propagator goes to the same massless form as in GR in the limit, the factor 1/6 in \eqref{7} will make sure that this term will give exactly the Van Dam-Veltman discontinuity. However, there is still an awkward $\epsilon$ there which comes from the gauge fixing. But it will be canceled after combined with the gauge fixed graviton propagator. Henceforth we will drop the tilde of $h_{\mu\nu}$. The graviton sector after full gauge fixing is,
\begin{eqnarray}
\frac{1}{4}({h_{\mu\nu}}^{,\lambda}{h^{\mu\nu}}_{,\lambda}-h^{,\lambda}h_{,\lambda}-2{h^{\mu\lambda}}_{,\lambda}{h_{\mu\lambda}}^{,\lambda}+2h^{,\mu}{h_{\mu\lambda}}^{,\lambda})-{\epsilon \over 6(1-\epsilon)}({h^{\mu\nu}}_{,\nu}-h,\mu)^2\nonumber \\
-{m_g}^2(h_{\mu\nu}h^{\mu\nu}-h^2)+\chi^2 h^2
 \end{eqnarray}
At this moment we are not going to consider the $\chi$ dependence since it will disappear in those terms that we are interested in when mass $m_g$ goes to 0. With this kinetic term, we obtain after some calculation the graviton propagator with a term that depends on $\epsilon$ and does not vanish in the massless limit. This tree-level contribution from this term to the graviton propagator is,
\begin{equation}
-({1-1/4 D \over 3/4 D-2 } -1/2) {\eta^{\mu\nu}\eta_{\alpha\beta} \over p^2}= +{\epsilon \over 6}{\eta^{\mu\nu}\eta_{\alpha\beta} \over p^2}.
\end{equation}
This will lead to an extra gauge dependent interaction between two sources as 
\begin{equation}+{\epsilon \over 6}{T^{a} T^{b} \over p^2}\end{equation}
which will exactly cancel the gauge dependent part in \eqref{7}. So the amplitude is indeed gauge independent.\\

\section{Generalization of the Fierz-Pauli Term to a Non-Unitary Gauge}
In fact only when the mass term takes the Fierz-Pauli form \eqref{5}, one will have such a discontinuity. For other kinds of mass terms, the gauge fixing term used to remove the mixing between $\pi_\mu$ and $h_{\mu\nu}$ is similar to \eqref{3a} but does not have the extra symmetry and the kinetic matrix is invertible. So we don't really need the redefinition and can get a propagator that returns to the same form as in GR in the massless limit. The Fierz-Pauli mass term is so special that one might expect it to be the  "unitary gauge" of some extra symmetry and the mass term will take a more general form in other gauges. Here we construct such a linearized theory with an extra local symmetry.\\
We consider the massive graviton Lagrangian under the following transformation.
\begin{equation}\label{Weyl} g_{\mu\nu} \to e^{2\omega} g_{\mu\nu}\end{equation} 
To the first order, this is 
\begin{equation}\label{Weyl2}h_{\mu\nu} \to h_{\mu\nu}+2\omega \eta_{\mu\nu}\end{equation} 
We already know that \eqref{2} will give 
\begin{equation}{M_{Pl}}^2 (\omega_{,\mu,\nu}h^{\mu\nu}-h\Box \omega)\end{equation} 
under \eqref{Weyl2}.  This can be compensated by the following gauge transformation of $\pi_\mu$.
\begin{equation}\pi_\mu \to \pi_\mu - 2{\omega_{,\mu} \over {m_g}^2 }\end{equation} 
But the Fierz-Pauli mass term, given by the second term of Eq. (\ref{Lag}) with $a=-b=f^4$, under this transformation \eqref{Weyl2} will also generate the following term,
\begin{equation}\label{31} \frac{f^4}{2} (-6 \omega H)=-3 \omega f^4 H \end{equation} 
If we introduce a new field $\psi$ with an interaction term
\begin{equation}3 \psi f^4 H \end{equation} 
then the transformation $\psi \to \psi + \omega$ will cancel \eqref{31}. However, in order to cancel the change of this new term under \eqref{Weyl2}, which is
\begin{equation}24 \omega f^4 \psi + 12 {M_{Pl}}^2 {\omega_{,\mu}}^{,\mu} \psi \end{equation}  
we need two more terms,
\begin{equation}-12 f^4 \psi ,\end{equation} 
and
\begin{equation} 6{M_{Pl}}^2 \psi_{,\mu}\psi^{,\mu}.\end{equation} 
In a word, to construct a new theory, we add the following terms to the Lagrangian,
\begin{equation}\label{psiHmix}3 \psi f^4 H-12 f^4 \psi^2 +6{M_{Pl}}^2 \psi_{,\mu}\psi^{,\mu}\end{equation} 
Since the $h_{\mu\nu}$ and $\pi_\mu$ are combined to form $H_{\mu\nu}$, these terms are still gauge invariant under the infinitesimal coordinate transformation \eqref{gauge}. Since we have this new symmetry, we need an extra gauge fixing term. In order to remove the mixing of H and $\psi$, we choose this gauge fixing term to be
\begin{equation}\label{GF2} \frac{3}{2} f^4 {(\beta H - {\psi \over \beta})}^2\end{equation} 
After canceling the mixing term of Eq. (\ref{psiHmix}), it will give two terms,
\begin{equation}\frac{3}{2} f^4 \beta^2 H^2 + \frac{3}{2} f^4 {\psi^2 \over \beta^2} \end{equation} 
In order to be invariant under \eqref{Weyl}, the coupling of $\psi$ and matter must be $e^{-2\psi}g_{\mu\nu}T^{\mu\nu}$. To linear order, this coupling is $(1 -2\psi)(\eta_{\mu\nu}+h_{\mu\nu})T^{\mu\nu}$. Let us define $\psi' = 2\psi$ in order to put this coupling term into the same canonical form as it had with $\phi_C$ in the previous section. The $\psi'$ terms coming from the sum of Eqs. (\ref{psiHmix}) and (\ref{GF2}), are
\begin{equation}\label{Psi-sector} \frac{3}{2} {M_{Pl}}^2 ({\psi'}_{,\mu}{\psi'}^{,\mu}+2 {m_g}^2 {\psi'}^2 - {{m_g}^2 \over 4 \beta^2} {\psi'}^2)\end{equation} 
One can see immediately that this scalar sector of $\psi'$ has the same form of kinetic term as $\phi$ had in \eqref{scalar-K} with $\epsilon=0$. Thus it will also give a 1/6 contribution in the massless limit. On the other hand, the gauge fixing term \eqref{GF2} will give a term proportional to $H^2$ and make the mass term no longer Fierz-Pauli like. (See the last term in Eq. (\ref{G-sector}).)

But, as in the previous Section, the Fierz-Pauli mass term is the only one that does not have a continuous limit that approaches the massless graviton propagator. Thus, the graviton propagator in this theory will return to the GR form in massless limit. But the conclusion regarding the vDVZ discontinuity is unchanged. We still have the discontinuity as a result of the scalar field contribution, even though we no longer have the Fierz-Pauli mass term. In the limit that $\beta \to 0$, one regains the Fierz-Pauli mass term and at the same time removes the $\psi'$ field by sending its mass to infinity. This physically unchanged behavior is what one would expect if the present massive gravition theory with a non-Fierz-Pauli Lagrangian is the manifestation of a fully nonlinear theory in a non-unitary gauge.

Finally let's check that this theory indeed gives a gauge-independent result, namely that $\beta$ will drop out from our final result. The massive graviton sector of this theory is 
\begin{eqnarray}
\label{G-sector}
\frac{1}{4}({h_{\mu\nu}}^{,\lambda}{h^{\mu\nu}}_{,\lambda}-h^{,\lambda}h_{,\lambda}-2{h^{\mu\lambda}}_{,\lambda}{h_{\mu\lambda}}^{,\lambda}+2h^{,\mu}{h_{\mu\lambda}}^{,\lambda})-{\frac{1}{2} \zeta^2}({h^{\mu\nu}}_{,\nu}-h,\mu)^2\nonumber \\
 -{m_g}^2(h_{\mu\nu}h^{\mu\nu}-(1-6\beta^2) h^2)
\end{eqnarray}
With some effort one can show that the graviton propagator from this action \eqref{G-sector} will contain the following term with the $\beta$ parameter,
\begin{equation}{{m_g}^2-6(k^2+{m_g}^2)\beta^2 {\eta^{\mu\nu}\eta_{\alpha\beta}} \over 3(k^2-{m_g}^2)(4k^2 \beta^2 + {m_g}^2 (-1+8 \beta^2))} \end{equation}
while the scalar $\psi'$ will give the contribution
\begin{equation}\frac{1}{6} {{\eta^{\mu\nu}\eta_{\alpha\beta}} \over k^2 - {m_g}^2(-2+{1\over 4 \beta^2})}\end{equation}
The sum of these two terms is 
\begin{equation}-\frac{1}{3} {{\eta^{\mu\nu}\eta_{\alpha\beta}} \over k^2 - {m_g}^2} \end{equation}
which is exactly what we want for the massive graviton without any gauge parameter $\beta$.\\

\section{Conclusion}
In Section 2, we calculated the graviton and scalar propagators for a 
gauge invariant massive graviton theory with Fierz-Pauli mass term and showed that their combined contribution is gauge independent. We showed that
in the limit that the graviton mass $m_g \rightarrow 0$, the 
contribution to the interaction of two sources, $T^{a}_{\mu\nu}$
and $T^{b}_{\mu\nu}$, differs from that of general relativity 
in momentum space by
an additional contribution of $\frac{1}{6 p^2} T^{a}T^{b}$, 
where $p$ is the $4$-momentum transfer.
This is exactly the vDVZ discontinuity that they obtained
in the gauge in which the Goldstone boson is absent, i.e., the
unitary gauge.
 
In Section 3, we introduced a new scalar field $\psi$ having
a mass proportional to $m_g$, such that there is an overall
local conformal symmetry of the theory. In this theory, we
introduce a class of gauge-fixing terms for which the
graviton mass term is different from that of Fierz and Pauli,
and in which the graviton propagator has a 
continuous massless limit. Nevertheless, we show that the
interaction between two external sources has the same
additional term as would arise from a Fierz-Pauli mass term.
Although there is no discontinuity in the massless limit
of the graviton propagator in this theory, the interaction 
of $\psi$ with the sources leads to the same extra
contribution found in Section 2. For a particular choice of the
gauge-fixing parameter, $\beta$, the theory reduces to that
given in Section 2. In the massless limit, the theory loses its 
local conformal symmetry and only retains a global Weyl symmetry.
It is not known if the present theory with local conformal
symmetry is the linearized form of a full nonlinear theory having
such a symmetry.
\\

\pagebreak

\end{document}